\documentstyle[eqsecnum,aps]{revtex}

\begin{document}
\draft
\title{Granular Collapse as a Percolation Transition}
\author{Jan Tobochnik\cite{byline}}
\address{Department of Physics, Kalamazoo College, Kalamazoo, MI 49006}
\date{\today}
\maketitle
\begin{abstract}Inelastic collapse is found in a two-dimensional
system of inelastic hard disks confined between two walls which act as an
energy source. As the coefficient of restitution is lowered, there
is a transition between a state containing small
collapsed clusters and a state dominated by a large collapsed cluster. The transition is
analogous to that of a percolation transition. At the transition the number of clusters
$n_s$ of size $s$ scales as $n_s \sim s^{-\tau}$ with $\tau \approx 2.7$.
\end{abstract}


There is much curent interest in developing a better
understanding of the behavior of granular systems such as sand,
powders, and model systems of metal and glass
beads~\cite{Jaeger}. These systems can behave as a solid, liquid,
or gas depending on the external conditions. However, granular systems behave very
differently than molecular systems and can exhibit size segregation, avalanches,
pattern formation, clustering, and collapse~\cite{TPC,GZ,Olafsen,MY,HL}. In
particular, clustering and collapse have been observed not just in
isolated dissipative systems, but also in driven
systems~\cite{Olafsen,Du,Kudrolli}, where energy is supplied from an
external source so that the system reaches a steady state. To
understand this phenomenon will likely require a number of
theoretical tools.

It is thus tempting to use the traditional tools of statistical physics~\cite{Jaeger2},
kinetic theory, and hydrodynamics~\cite{TPC,Haff,JR} to analyze the
behavior of these systems. For example, a hydrodynamic description
which stems from the Boltzmann equation depends on the assumption
that interparticle correlations do not exist~\cite{Grossman}. However, this work 
was confined to near elastic systems, and cannot explain 
the phenomena of inelastic collapse which can occur far from the elastic limit.
Although it is possible to define a temperature proportional to the mean kinetic
energy per particle, there is no thermal equilibrium equivalent to
that found for molecular systems. For example, simulations and
experiments have shown that granular systems do not follow a simple
Maxwell-Boltzmann velocity distribution, and there is no
equipartition of energy~\cite{GZ,Du,Grossman,LR,TT,Warr}. Hydrodynamic theories have
been able to predict the length scales of clustering in granular gases~\cite{Petz},
however no theory has yet been able to predict the behavior of collapsed systems. 
In this paper we look at the inelastic collapse
of a model driven granular system from the point of view of
percolation theory, which describes the transition from a configuration of
isolated small clusters to a configuration dominated by a single large
cluster which spans the system. We will make an analogy between the
percolation transition and a granular collapse transition. It is important to
distinguish the phenomena of clustering from that of collapse discussed here.
Clustering refers to density inhomogeneities where particles in a cluster have
correlated motion. Collapse refers to the case where particles have lost essentially
all their kinetic energy due to inelastic collisions. The novel aspect of our work is
that the collapse of our system occurs over a wide range in the degree of
inelastically, but the geometry of the collapsed particles appears to behave in a
manner similar to that of a percolation transition. 

To investigate the geometry of the collapsed disks in a driven system of
inelastic hard disks, we use event driven molecular dynamics simulations~\cite{GZ}. The
hard disks are confined to a simulation cell of size $L_x
\times L_y$ with periodic boundary conditions in the
$x$ direction and hard walls at $y = 0$ and $y = L_y$. The two walls
are held at a fixed wall temperature $T_w$. All
lengths are measured in units of the disk diameter. The system
reaches a steady state when the energy dissipated through the
collisions is compensated by energy supplied by the two walls. When
a particle collides with the wall, it is ejected with a velocity
whose components are distributed according to the probabilities:
\begin{equation}
\label{pv}
P(v_x) = {1 \over \sqrt{2\pi}} e^{-v_x^2/2T_w} \quad
\mbox{and}
\quad P(v_y) = v_y
e^{-v_y^2/2T_w}, 
\end{equation}
where it is understood that $v_y$ takes on only positive values for the bottom wall
and negative values for the top wall.   For elastic disks this probability distribution
leads to a Maxwell-Boltzmann(MB) probability distribution for the particles near the
wall. For elastic spheres it has been shown\cite{Tehver} that using the functional
dependence of
$P(v_x)$ for $P(v_y)$ (instead of using Eq.~(\ref{pv})) may lead to deviations from MB
velocity distributions, temperatures away from the wall not equal to $T_w$, and
unphysical inhomogeneities in density and temperature.   Actual experimental systems
which use vibrating walls will typically not give MB velocity distributions either.
However,  Grossman et al~\cite{Grossman} found that their results away from the wall
were not sensitive to the details of the boundary conditions. For our purposes the only
function of the wall is to input energy into the system so that the system can come to
a steady state with non-zero energy. The value of using boundary conditions which give
the correct MB velocity distribution for elastic disks is that we can test our program
in the limit of elastic disks.  In our simulations we set
$T_w = 1$. The specific value of $T_w$ merely sets the time scale in the problem. We
have performed simulations with different temperatures for each wall, but the results
are qualitatively similar to those found when both walls are at the same temperature. 

During a collision between two particles, $i$ and $j$, momentum is
conserved and some energy lost. We denote the
component of the velocity perpendicular to the line connecting the
center of the two disks by $v_{i \perp}$ and the parallel component
by
$v_{i \parallel}$. We denote velocities after the collision by a
prime. The amount of energy lost in a collision between two disks
depends on the coefficient of restitution,
\begin{equation}
R \equiv {|v'_{i \perp} - v'_{j \perp}| \over |v_{i \perp} -
v_{j \perp}|}
\end{equation} The case $R=1$
corresponds to the elastic limit. By momentum conservation
$v'_{i \parallel} = v_{i \parallel}$ and $v'_{j \parallel} =
v_{j \parallel}$. The velocities after the collision are given by
\begin{equation}
\left(\matrix{v'_{i \perp}\cr v'_{j \perp}\cr}\right) = {1 \over 2}
\left(\matrix{1-R&1+R\cr 1-R&1+R \cr}\right)
\left(\matrix{v_{i \perp}\cr v_{j \perp}\cr}\right)
\end{equation}
The mechanical energy lost in the collision becomes
\begin{equation}
E_{\rm lost} = -{1 \over 4}(1-R^2)(v_{i \perp} - v_{j \perp})^2.
\end{equation}

The initial configuration for the particles consists of a random
placement of the particles in the simulation cell, with velocities
distributed according to a Maxwell-Boltzmann distribution at
temperature $T_w$. Unless otherwise specified, the number of
particles is $N = 500$. Typically, the system has reached a steady
state after about
$10^5$ collisions, and the properties of the steady state are
analyzed. The time used in the simulation is set such that all macroscopic quantities
do not systematically change with time. The simulation is repeated for typically
100 runs to collect data.

We define a ``temperature''
$T$ as the mean kinetic energy per particle. The behavior for the
temperature , density $\rho$, and pressure $P$ as a function of $y$
is shown in Fig.~\ref{f1} for a $30 \times 30$ system of 500 disks. The resolution used
to make measurements is one disk diameter. The pressure can be obtained from the time
averaged impulse received by the particles during collisions. The pressure is found
to be uniform throughout the system as required by mechanical
stability. The density is smaller very close to the walls. This
behavior occurs more strongly in relatively dense systems, and diminishes as
the mean density of the system decreases. The temperature reaches a
minimum and the density a maximum in the center of the system for
all $R < 1$. This qualitative behavior is easily explained.
Particles can receive energy from the walls. As they move away from
the walls, energy is dissipated in collisions. By symmetry we would
expect the particles at the center to have the least energy and
thus the lowest temperature. Note that the temperature of the
particles whose centers are within a particle diameter of a wall is
significantly less than the wall temperature
$T_w = 1.0$. This behavior is due to the inability of the particles
to come to local thermal equilibrium. Fast moving particles, which
have just picked up energy from the wall and are moving away from
it, cannot thermalize with the slower particles moving toward the
wall. Thus, the mean kinetic energy of the particles near the walls
is an average over the higher energy particles moving away from the
wall and the lower energy particles moving toward the wall.  Only the
temperature of the particles moving away from the wall have a temperature
equal to that of the wall. If the temperatures
of the two walls are different, then our simulations show a
temperature minimum and density maximum that is shifted toward the
lower temperature wall.

Because particles at the center have less energy, there is a
tendency for these particles to pack together, and thus a density
maximum occurs. We would expect this kind of behavior even in the
ideal gas limit where
$\rho = P/T$.  Grossman et al.~\cite{Grossman} developed
an approximate hydrodynamic theory for their system, which worked
well for values of $R$ close to unity. We have adapted their theory
to our system and found that it can reproduce the
temperature minimum, but quantitative agreement is only very
approximate, even at values of $R$ near unity.  

For $R$ close to unity and a sufficiently low  density of particles, the system will
consist of a small number of slow moving particles,
$N_s$, defined as particles with a kinetic energy less than 0.01. The value of the
kinetic energy cutoff was chosen so that the cluster labeling discussed below is
insensitive to small changes in this value, and that the cutoff is low enough that
there would be very few particles emerging from the wall with a kinetic energy lower
than the cutoff.  As $R$ is lowered, the number of slow moving particles increases, and
some of the slow moving particles begin to cluster into groups of disks in a collapsed
state. This becomes evident because the mean distance moved by slow particles during a
collision is of order $10^{-4}$ of a disk diameter. Particles in these clusters
oscillate about an equilibrium position similar to molecules in a crystal. Even though
energy is lost  on each collision, there is energy pumped in from the surface of the
collapsed clusters due to faster moving particles which are not part of the collapsed
cluster. Thus, the kinetic energy of particles in the collapsed clusters does not
vanish. However, these collapsed clusters do not occur for the same reason as elastic
hard disk(or sphere) solidification. In the latter case hard disks can have any amount
of kinetic energy, because the kinetic energy only determines the time scale of the
simulation. Solidification occurs because of an imposed high density which restricts
the motion of the disks. In the granular case here, the collapsed state occurs because
the disks lose almost all their kinetic energy. The mean density of our simulation cell
is well below the freezing density for hard disks.  

Two nearby slow particles are defined to be in the same cluster if they are
separated by a distance less than 1.05. This cutoff definition was chosen so that small
changes in its value do not change the definition of the clusters. A continuum version
of the Hoshen-Kopelman algorithm\cite{HK} is used to define the clusters. Lowering
$R$ leads to clusters of larger size, and eventually to a state with one large
collapsed cluster and several very small clusters. At this point, typically the second
largest cluster is less than one fifth the size of the largest cluster. The largest
cluster has a stable hexagonal crystal structure with only its surface changing with
time during the time scale of our simulations. Thus, the particles can be divided into
two groups: those that are part of collapsed solid-like clusters of particles which are
moving very slowly, and a dilute collection of uncorrelated particles which are moving
quickly and thus constitute a granular gas. 
 Fig.~\ref{f2} shows snapshots of two systems which contain a large solid-like cluster 
surrounded by smaller clusters and isolated disks. These snapshots are typical of what
is seen near the transition. As $R$ is lowered the solid-like largest cluster grows and
its hexagonal crystal structure contains fewer defects. 

The above scenario is analogous to that found in standard
percolation theory, where cluster properties are
calculated as a function of a parameter $p$ which is the fraction of
occupied sites (for a lattice model) or a volume fraction of the
space covered by the objects of interest. At
$p = p_c$, one large cluster forms which spans the entire system.
The percolation threshold becomes more sharply defined as the size
of the system increases. In the infinite size limit, the
connectedness length $\xi$, which is a measure of the linear
dimension of the non-spanning clusters, diverges as $\xi \sim
|p-p_c|^{-\nu}$. The mean size of the non-spanning clusters, $\chi$
diverge as $\chi \sim |p-p_c|^{-\gamma}$. The fraction of the
occupied space which is part of the spanning cluster,
$P_{\rm span}$ vanishes as $(p-p_c)^{\beta}$, where $\beta$ is
another critical exponent. In addition at $p_c$ the number of
clusters of size $s$ per lattice site,
$n_s$, scales as $n_s \sim s^{-\tau}$. Scaling theory leads to the 
relations $2\beta +
\gamma =
\nu d$ and $\tau = 2 + \beta/(\beta + \gamma)$~\cite{Stauffer}. The
exponents $\nu$, $\gamma$, $\beta$, and $\tau$ depend only on the
dimensionality
$d$ of space.

Percolation theory has been very useful for describing a large variety of
randomly disordered systems. The theory provides a guide for determining which
quantities show power law behavior near a transition. In addition the theory makes many
universal predictions which are independent of the details of how the disorder is
created. For these reasons, we believe it is useful to describe granular collapse in
our system with percolation theory as a guide. 

To make the analogy with percolation theory, we define $p = N_s/N$.
At $p=p_c$ we expect to see collapsed clusters of all
sizes in the limit as $N \to \infty$. In analogy to percolation theory we define a
spanning cluster as a cluster which spans the simulation cell in the $x$ direction.    
 Explicit formulae for calculating
the percolation quantities are as follows. The connectedness length is given by:
\begin{equation}
\xi^2 = {1 \over 2} { \sum_{k= 1}^{N_c} \sum_i \sum_j r_{i,j,k}^2 \over
\sum_{k= 1}^{N_c} s_k^2} ,
\end{equation}
where $N_c$ is the number of collapsed clusters not including the spanning
cluster, $r_{i,j,k}$ is the distance between disk $i$ and disk $j$
in cluster $k$, and $s_k$ is the number of disks in the $k$th
cluster. The mean cluster size is
\begin{equation}
\chi = { \sum_{k= 1}^{N_c} s_k^2 \over \sum_{k= 1}^{N_c}
s_k}
\end{equation}
Another useful quantity is the fraction of runs which contain a
spanning cluster which we denote by $f$. We define $P_{\rm span}$ as
the number of particles in the spanning cluster divided by $N_s$.
 
Figure~\ref{f3} shows how $p$ depends on $R$ for a $30
\times 30$ simulation cell with $N=500$ particles. Because the
dependence is smooth over the entire range of $R$, we expect
any power law behavior to be the same as a function of $R$ or $p$.
 Figure~\ref{f4} shows the
results for the quantities
$f$, $P_{\rm span}$,
$\xi$, and
$\chi$. These results are qualitatively similar to those found in
typical continuum percolation systems. Our system sizes are much too small to be
able to extract reliable critical exponents or a precise value of
$p_c$. However, if we define $p_c$ as the value of $p$ where $f =
1/2$ and make crude estimates for the critical exponents, we find
$\nu
\approx 1$, $\gamma
\approx 1.5$, and $\beta
\approx 0.5$. The estimates for $\nu$ and $\gamma$ are approximately the same
above and below the transition, a result expected from percolation theory. 
Our values lead to $2 \beta + \gamma = 2.5$ and
$2\nu = 2$. Because our
estimates are very crude the scaling relation cannot be ruled out. For
comparison the values found in standard 2D percolation are
$\nu = 4/3$,
$\gamma = 43/18$, and
$\beta = 5/36$. Much better statistics
 with a larger number of disks would be needed to  estimate the percolation exponents
reliably and determine if scaling exists.

Near $p_c$, the power law dependence of the cluster size
distribution is robust and always shows a critical exponent of
$\tau
\approx 2.7$ for several different values of $N$ and cell size,
including asymmetric simulation cells such as a $20 \times 160$
cell. Figure~\ref{f5} shows typical log-log plots of $n_s$ versus
$s$. For $N=1000$ we have a sufficient number of runs
and number of disks to obtain statistics on clusters up to size
$s = 30$. For the other systems the plots show data for clusters up
to size
$s=10$. This estimate for $\tau$ is much different than that found
in standard 2D percolation where $\tau = 187/91 \approx 2.06$. If
our result for $\tau$ is approximately correct, then it is very
difficult to see how the scaling result $\tau = 2 + \beta/(\beta +
\gamma)$ could be valid. One possibility is that the relationship
between cluster numbers and the other percolation quantities is not
the same as that found in standard percolation theory. This could be because in our
case there is a systematic variation in density from one wall to the
other in the y direction, and for $L_y$ large enough our results are independent of
$L_y$ once a steady state has been reached. Thus, no cluster properties are normalized
by the linear dimensions of the system. On the other hand standard percolation systems
do not have this systematic inhomogeneity, and cluster numbers are normalized by the
linear dimensions of the system. Another possibility is that our estimates for the
critical exponents are too crude. 

Many of the features we observe are similar to those found in percolation theory.
The connectedness length and mean cluster size grow very quickly with small changes in
the coefficient of restitution near the transition in analogy to how these quantities
grow quickly as a function of an occupation factor $p$ in percolation systems. There
are large variations from run to run in percolation quantities near the transition
suggestive of the the critical fluctuations one finds near a standard percolation
transition or any second order phase transition. These critical fluctuations are due
to the presense of structures of all length scales. Our robust power law behavior for
the cluster size distribution provides further evidence for a true phase transition in
the limit $N \to \infty$.  

In summary, we have described the inelastic
collapse of a model system of granular material in terms of
percolation type quantities. We have found that the number of
clusters of size $s$ behaves as a power law, and that other
quantities behave in a way similar to analogous quantities in
percolation theory. Much larger systems are needed to find
quantitatively precise values of the exponents analogous to
critical percolation exponents. These results suggest that the geometry of 
collapsed clusters may be an important universal feature of collapse in
driven granular systems. 

I am grateful to Arshad Kudrolli for many useful discussions and
Christopher Grostic for catching an error in my original program. I
acknowledge support from the donors of the Petroleum Research Fund
administered by the American Chemical Society.

\begin{figure}
\caption{The temperature $T$, pressure $P$, and density $\rho$ as a
function of position from the wall at $y = 0$.}
 \label{f1}
 \end{figure}

\begin{figure}
\caption{Snapshots of two systems which contain a solid-like cluster.
The parameters are (a) number of particles $N = 2000$, linear
dimensions $L_x = 80$,
$L_y = 160$, and coefficient of restitution
$R = 0.89$; (b) $N = 500$, $L_x = L_y = 30$, $R = 0.85$. }
 \label{f2}
 \end{figure}

\begin{figure}
\caption{The fraction of slow particles, $p$, as a function of  $R$.
with $N = 500$, $L_x = L_y = 30$, averaged over 100 runs.}
 \label{f3}
 \end{figure}

\begin{figure}
\caption{Percolation properties as a function of
the fraction of slow particles, $p$. (a) Plot of $f$, the fraction
of runs which contain a spanning cluster, and $P_{\rm span}$, the
fraction of slow particles in the spanning collapsed cluster; (b) the mean
cluster size $\chi$, and (c) the connectedness length
$\xi$ in units of the disk diameter. Properties computed from the same system as in
Fig.~\ref{f3}. The statistical uncertainties of $\chi$ and $\xi$ are approximately
$10\%$. The curves are only guides to the eye.}
\label{f4}
\end{figure}

\begin{figure}
\caption{Log-log plot of the mean number of clusters $n_s$ of size $s$
versus $s$ for five different systems. The lines are least squares fits
for each system. The average of the slopes is 2.7. }
 \label{f5}
 \end{figure}

%
%

\end{document}